\documentclass[11pt]{article}
\usepackage{moriond,epsfig}
\usepackage{amsmath}

\def\be{\begin{equation}}
\def\ee{\end{equation}}
\def\bea{\begin{eqnarray}}
\def\eea{\end{eqnarray}}

\begin{document}
\vspace*{4cm}
\title{\boldmath CP Violation in $D^0 - \bar D^0$ Mixing and Electric Dipole Moments \\ in SUSY Alignment Models\footnote{talk given at Rencontres de Moriond EW 2011, March 13 - 20, 2011}}

\author{W. ALTMANNSHOFER}

\address{Theoretical Physics Department, \\ Fermilab, P.O. Box 500, Batavia, IL 60510, USA}

\maketitle\abstracts{
We report on a study of CP Violation in $D^0 - \bar D^0$ mixing and Electric Dipole Moments in the framework of supersymmetric alignment models. Both classes of observables are strongly suppressed in the Standard Model and highly sensitive to new sources of flavor and CP violation that can be present in models of New Physics. 
Supersymmetric alignment models generically predict large non-standard effects in $D^0 - \bar D^0$ mixing and we show that visible CP violation in $D^0 - \bar D^0$ mixing implies lower bounds for the EDMs of hadronic systems, like the neutron EDM and the mercury EDM, in the reach of future experimental sensitivities.
We also give updated constraints on the mass insertions of the Minimal Supersymmetric Standard Model using the current data on $D^0 - \bar D^0$ mixing.}

\section{Introduction}

Models of New Physics (NP) often contain new sources of flavor violation and are therefore strongly constrained by experimental data on Flavor Changing Neutral Current (FCNC) processes. This is in particular the case for Supersymmetric (SUSY) extensions of the Standard Model (SM) like the Minimal Supersymmetric Standard Model (MSSM), as long as the SUSY degrees of freedom are not far above the TeV scale~\cite{Altmannshofer:2009ne,Gabbiani:1996hi}.
This so-called SUSY flavor problem is for example addressed in SUSY alignment models~\cite{Nir:1993mx,Leurer:1993gy} that align the down quark and down squark mass matrices such that down quark -- down squark -- gaugino couplings are flavor diagonal and FCNC processes in the down sector are under control. A characteristic prediction of these models are however sizable NP effects in up sector FCNCs, in particular in $D^0 - \bar D^0$ mixing~\cite{Nir:1993mx}.
On general grounds $D^0 - \bar D^0$ mixing observables are highly sensitive probes of the flavor sector of NP models~\cite{Bianco:2003vb}. Especially CP violation in $D^0 - \bar D^0$ mixing is strongly suppressed in the SM by O$(V_{ub}V_{cb}/V_{us}V_{cs}) \sim 10^{-3}$ and experimental evidence for it considerably above the per mill level would clearly point towards the presence of NP (see however \cite{Bobrowski:2010xg}).

In the following we give updated bounds on the mass insertions of the MSSM using the latest experimental data on $D^0 - \bar D^0$ mixing, we analyze the predictions of SUSY alignment models for CP violation in $D^0 - \bar D^0$ mixing and show that sizable CP violating effects in $D^0 - \bar D^0$ mixing imply lower bounds on the Electric Dipole Moments (EDMs) of hadronic systems within this class of SUSY models. The presentation is mainly based on~\cite{Altmannshofer:2010ad}.

\section{\boldmath Bounds on Mass Insertions from $D^0 - \bar D^0$ Mixing} \label{sec:Dmix}

The neutral $D$ meson mass eigenstates $D_1$ and $D_2$ are linear combinations of the strong interaction eigenstates, $D^0$ and $\bar D^0$
\begin{equation}
| D_{1,2} \rangle = p | D^0 \rangle \pm q | \bar D^0 \rangle ~,~~~ \frac{q}{p} = \sqrt{\frac{M_{12}^* - \frac{i}{2} \Gamma_{12}^*}{M_{12} - \frac{i}{2} \Gamma_{12}}} ~,
\end{equation}
where $M_{12}$ is the dispersive part and $\Gamma_{12}$ the absorptive part of the $D^0 - \bar D^0$ mixing amplitude
\begin{equation}
\langle D^0 |\mathcal{H}_{\rm eff}| \bar D^0 \rangle = M_{12} - \frac{i}{2} \Gamma_{12} ~,~~~\langle \bar D^0 |\mathcal{H}_{\rm eff}| D^0 \rangle = M_{12}^* - \frac{i}{2} \Gamma_{12}^* ~.
\end{equation}
The normalized mass and width differences in the $D^0 - \bar D^0$ system, $x$ and $y$, are given by
\begin{equation}
x = \frac{\Delta M_D}{\Gamma_D} = 2 \tau_D \textnormal{Re}\left[ \frac{q}{p} \left( M_{12} - \frac{i}{2} \Gamma_{12} \right) \right] ~,~~~
y = \frac{\Delta \Gamma}{2 \Gamma_D} = - 2 \tau_D \textnormal{Im}\left[ \frac{q}{p} \left( M_{12} - \frac{i}{2} \Gamma_{12} \right) \right] ~,
\end{equation}
with the lifetime of the neutral $D$ mesons $\tau_D = 1/\Gamma_D = 0.41$ps.

Experimentally, $D^0 - \bar D^0$ mixing is firmly established with the non-mixing hypothesis $x=y=0$ excluded at $10.2\sigma$~\cite{Asner:2010qj}. Still, at the current level of sensitivity, there is no evidence for CP violation in
$D^0-\bar D^0$ mixing. The experimental data on both $|q/p|$ and $\phi=\textnormal{Arg}(q/p)$ is compatible with CP conservation, i.e. $|q/p|=1$ and $\phi=0$. The most recent world averages as obtained by HFAG read~\cite{Asner:2010qj}
\begin{equation} \label{eq:Dmix_bounds}
x = (0.63^{+0.19}_{-0.20})\% ~,~~~ y = (0.75 \pm 0.12)\% ~,~~~ \left| q/p \right| = 0.91^{+0.18}_{-0.16} ~,~~~ \phi = (-10.2^{+9.4}_{-8.9})^\circ ~.
\end{equation}
These experimental results on $D^0 - \bar D^0$ mixing lead to strong constraints on possible new sources of flavor violation in extensions of the Standard Model~\cite{Ciuchini:2007cw,DDbar_NP,Altmannshofer:2009ne}. 

The MSSM contains many new sources of flavor violation. A convenient parametrization is given by so-called mass insertions $\delta$ that can be defined as the deviations of the up and down squark mass matrices from universality in the super-CKM basis
\begin{equation}
M^2_{\tilde q} = \tilde m^2_Q (1\!\!1 + \delta_q) ~~,~~~ \delta_q = \begin{pmatrix} \delta_q^{LL} & \delta_q^{LR} \\ \delta_q^{RL} & \delta_q^{RR} \end{pmatrix}  ~~,~~~ q = u,d ~~.
\end{equation}
Complex flavor off-diagonal mass insertions lead to flavor and CP violating gluino -- squark -- quark interactions that typically lead to huge contributions to FCNC processes.
Taking into account only gluino box contributions in the so-called mass insertion approximation, neglecting for simplicity renormalization group effects as well as setting the $B$-parameter to 1 in the evaluation of the hadronic matrix elements, one finds for the MSSM contributions to the $D^0 - \bar D^0$ mixing amplitude the following approximate expression~\footnote{In our numerical analysis we implement the full set of 1 loop MSSM contributions that can be found e.g. in~\cite{Altmannshofer:2007cs}, we include 2 loop renormalization group running~\cite{RGE2loop} and use the hadronic matrix elements given in~\cite{Ciuchini:2007cw}.}
\begin{equation} \label{MD_12}
M_{12}^{\rm NP} \simeq \frac{\alpha_s^2}{\tilde m_Q^2} m_Df_D^2 \left[ \Big( (\delta_u^{LL})_{12}^2 + (\delta_u^{RR})_{12}^2 \Big) \frac{g_1(x_g)}{3} + (\delta_u^{LL})_{12}(\delta_u^{RR})_{12} ~\frac{m_D^2}{m_c^2} \left( \frac{g_4(x_g)}{4} + \frac{ g_5(x_g)}{12} \right) \right] ~,
\end{equation}
where $m_D$ is the mass and $f_D$ the decay constant of the neutral $D$ mesons. The loop functions $g_1$, $g_4$ and $g_5$ depend on the ratio $x_g = M_{\tilde g}^2/\tilde m_Q^2$ of the gluino and squark masses and their explicit expression can be found e.g. in~\cite{Altmannshofer:2009ne}. In the limiting case of degenerate masses one has $g_1(1)=-\frac{1}{216}$, $g_4(1)=\frac{23}{180}$ and $g_5(1)=-\frac{7}{540}$. In~(\ref{MD_12}) we neglected contributions from $\delta_u^{LR}$ and $\delta_u^{RL}$ mass insertions. They are given e.g. in~\cite{Altmannshofer:2009ne}. 

\begin{figure} \centering
\psfig{figure=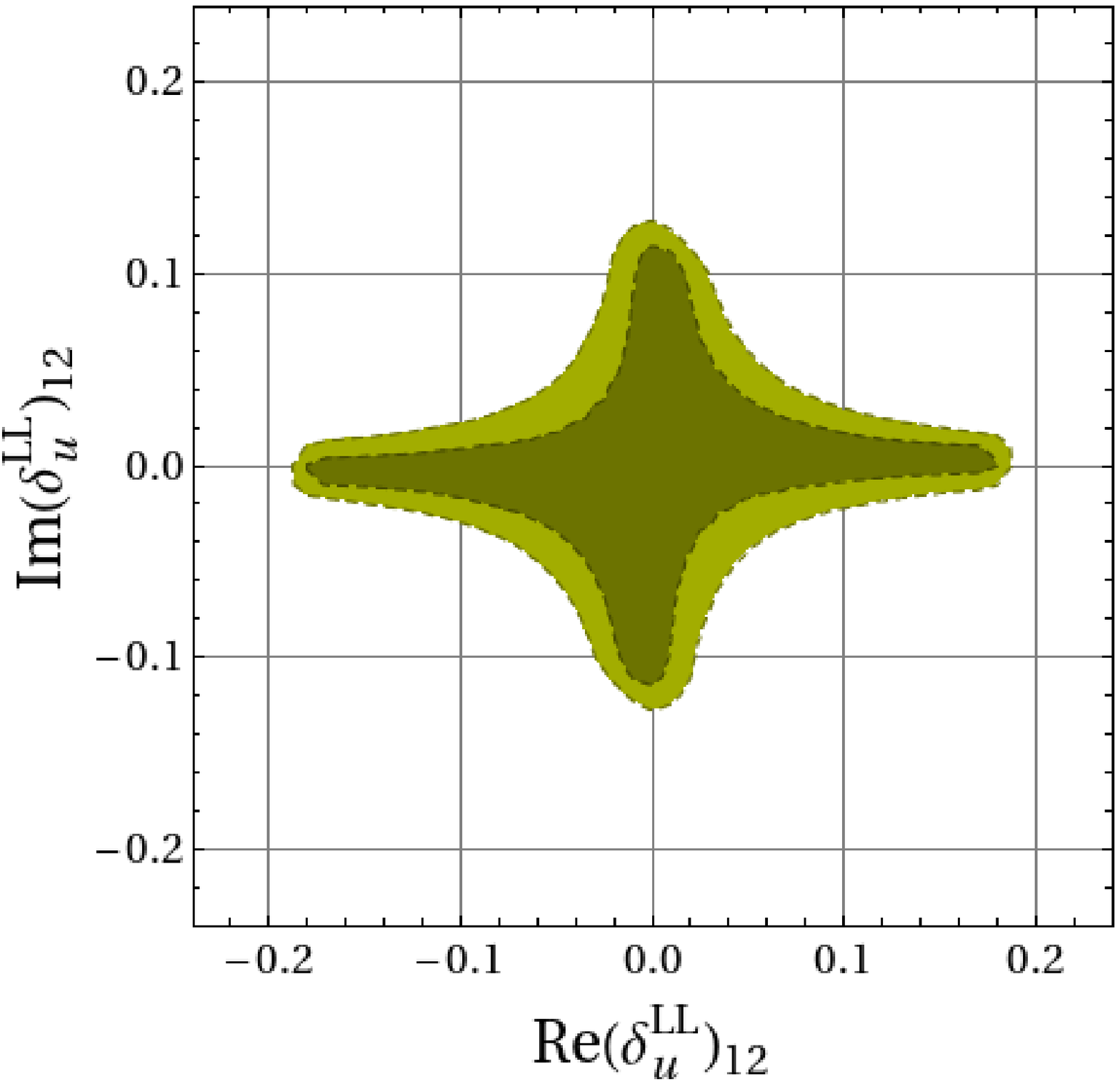,height=1.7in} ~~~~
\psfig{figure=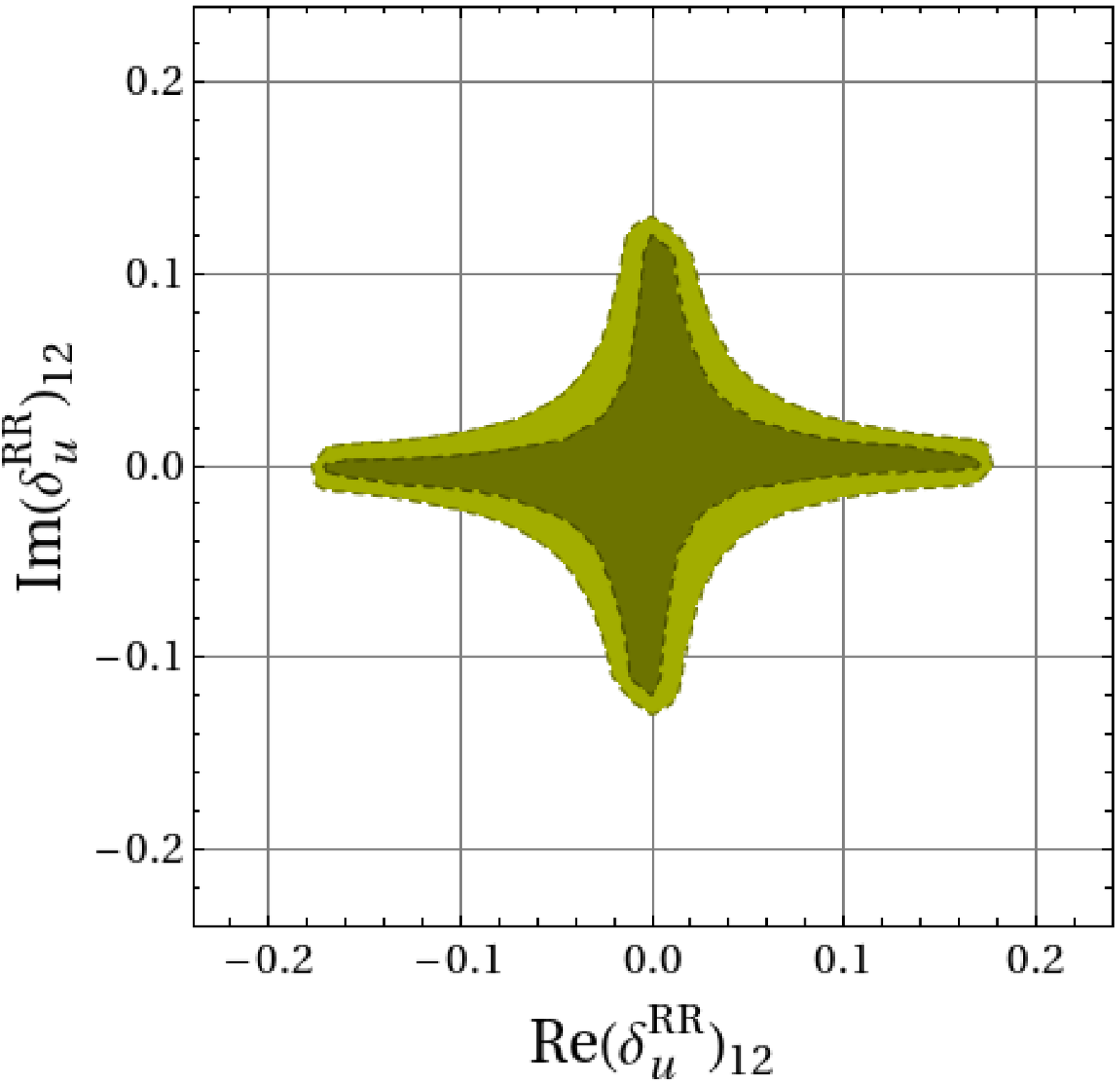,height=1.7in} ~~~~
\psfig{figure=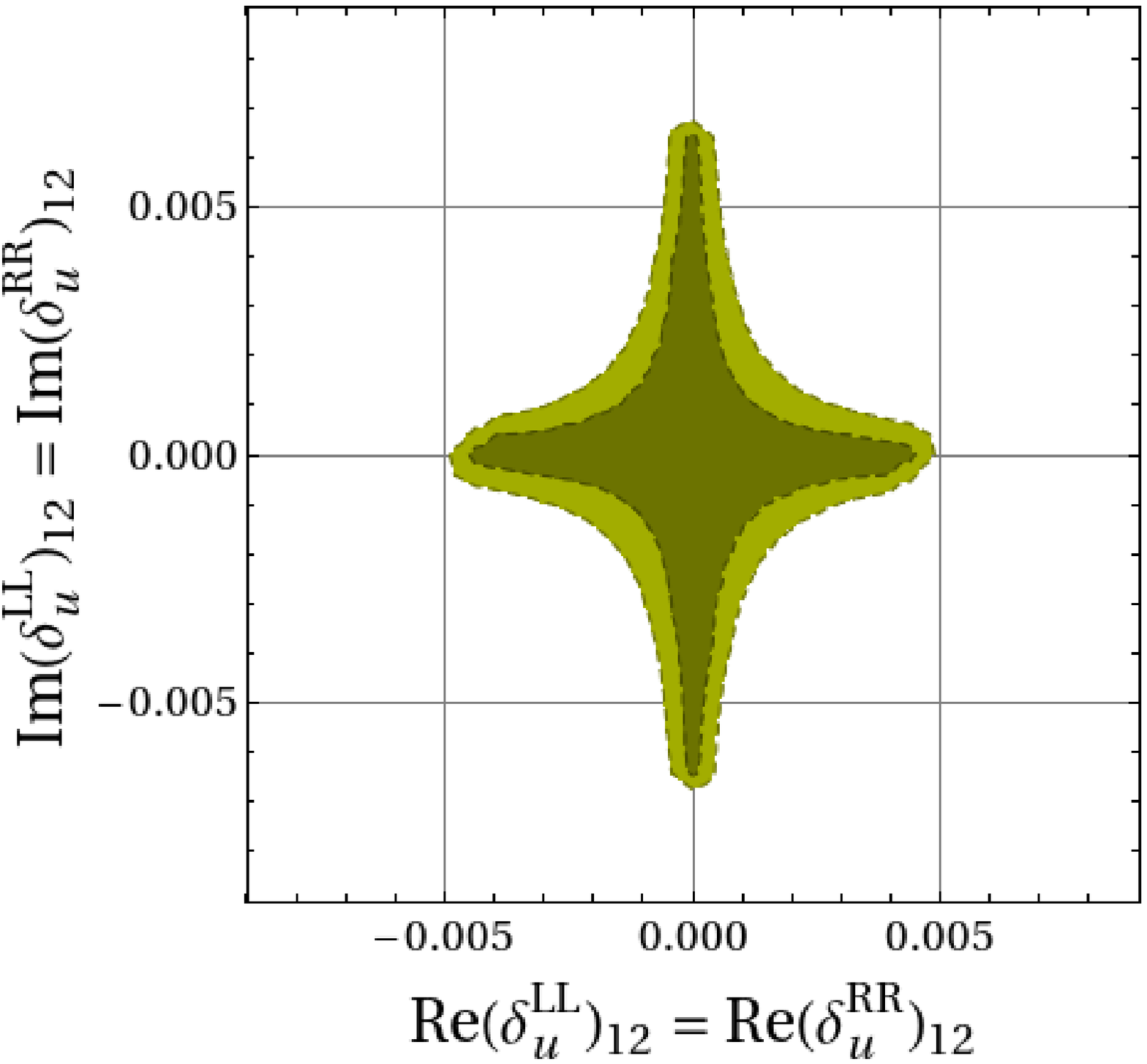,height=1.7in}
\caption{Updated constraints on the mass insertions $(\delta_u^{LL})_{12}$ and $(\delta_u^{RR})_{12}$ from $D^0 - \bar D^0$ mixing for a common squark and gluino mass of $\tilde m_Q = M_{\tilde g} = 1$~TeV.
\label{fig:deltas}}
\end{figure}

In Fig.~\ref{fig:deltas} we show the allowed regions for the mass insertions $(\delta_u^{LL})_{12}$ and $(\delta_u^{RR})_{12}$. As the SM contributions to $M_{12}$ and $\Gamma_{12}$ cannot be predicted in a reliable way, we allow them to vary in the range $-0.02$ps$^{-1} < M_{12}^{\rm SM} < 0.02$ps$^{-1}$ and $-0.04$ps$^{-1} < \Gamma_{12}^{\rm SM} < 0.04$ps$^{-1}$ and impose the constraints~(\ref{eq:Dmix_bounds}) at the $2\sigma$ level. The bounds on the mass insertions are obtained for a SUSY spectrum with a common squark and gluino mass of $\tilde m_Q = M_{\tilde g} = M_{\rm SUSY} = 1$TeV and switching on one mass insertion at a time. They scale as $\delta_u / M_{\rm SUSY}$ and hold barring accidental cancellations among the different contributions in~(\ref{MD_12}).

The case where both LL and RR mass insertions are present simultaneously is particularly strong constrained (see right plot of Fig.~\ref{fig:deltas}). Even for the rather heavy SUSY spectrum that we consider, the mass insertions have to be smaller then about $5 \cdot 10^{-3}$. For maximal phases of the mass insertions, the bounds are stronger by approximately a factor of 3.

\section{\boldmath CP Violation in $D^0 -\bar D^0$ Mixing in SUSY Alignment Models} \label{sec:CPV_Dmix}

A popular class of SUSY models that generically predict large NP effects in $D^0 - \bar D^0$ mixing are SUSY alignment models~\cite{Nir:1993mx}.
The quark-squark alignment mechanism occurs naturally in models with abelian horizontal symmetries that reproduce the
observed hierarchy in the SM Yukawa couplings. Interestingly, in the framework of alignment it is possible to predict for a broad class of abelian flavor models both lower and upper bounds for the mass insertions~\cite{Nir:2002ah}.

The most characteristic prediction of alignment models is the appearance of a large $(\delta_u^{LL})_{12}$ mass insertion that  leads to large effects in $D^0 - \bar D^0$ mixing. 
Indeed, $SU(2)$ invariance implies a relation between the left-left mass insertions in the up and down sector
\begin{equation}
(\delta_u^{LL})_{12} = (V \delta_d^{LL} V^\dagger)_{12} \simeq (\delta_d^{LL})_{12} + \lambda \frac{m^2_{\tilde c_L} - m^2_{\tilde u_L}}{\tilde m^2_Q} + O(\lambda^2) ~,
\end{equation}
where $V$ is the CKM matrix, $\lambda \simeq 0.2$ is the Cabibbo angle and $m_{\tilde u_L}$ and $m_{\tilde c_L}$ are the left handed up and charm squark masses, respectively. As abelian flavor symmetries do not impose any restriction on the mass splittings between squarks of different generations, they are expected to be non-degenerate with natural order one mass splittings.
Correspondingly, even for $(\delta_d^{LL})_{12} = 0$, which is approximately satisfied in alignment models to avoid the strong constraints from Kaon mixing, there is an irreducible flavor violating term of order $\lambda$ leading to $c-u$ transitions. Note that this $(\delta_u^{LL})_{12}$ is real to a good approximation.

As shown in~\cite{Nir:2002ah}, the right-right mass insertion leading to $c-u$ transitions is predicted to be $\lambda^2~<~|(\delta_u^{RR})|_{12}~<~\lambda^4$ in abelian flavor models with alignment. This mass insertion is naturally expected to be complex. Therefore, all CP violating phenomena in $D^0 - \bar D^0$ mixing are dominantly generated by the following combination of mass insertions
\begin{equation} \label{ImMD_12}
{\rm Im}M_{12}^{\rm NP} \propto {\rm Im}[(\delta_u^{LL})_{12}(\delta_u^{RR})_{12}]~.
\end{equation}
In the following we focus on two observables sensitive to CP violation in $D^0 - \bar D^0$ mixing: (i)~the semileptonic asymmetry $a_{SL}$ and (ii)~the time dependent CP asymmetry in decays to CP eigenstates $S_f$.

The semileptonic asymmetry in the decay to ``wrong sign'' leptons is defined as
\begin{equation}
a_{\rm SL} = \frac{\Gamma(D^0 \to K^+ \ell^- \nu) - \Gamma(\bar D^0 \to K^- \ell^+ \nu)}{\Gamma(D^0 \to K^+ \ell^- \nu) + \Gamma(\bar D^0 \to K^- \ell^+ \nu)} = \frac{|q|^4-|p|^4}{|q|^4+|p|^4}
\end{equation}
and is a direct measure of CP violation in the mixing. However, as the decay rates to the ``wrong sign'' leptons are strongly suppressed by $x^2 + y^2$, measurements of this asymmetry are experimentally challenging.

Also the time dependent CP asymmetry $S_f$ in decays to a common CP eigenstate $f$, aka lifetime CP asymmetry $\Delta Y_f$, is a sensitive probe of CP violation in $D^0 - \bar D^0$ mixing~\cite{Bergmann:2000id,Bigi:2009df}
\begin{equation} \label{eq:DYf_def}
S_f = 2 \Delta Y_f = \frac{1}{\Gamma_D} \left( \hat \Gamma_{\bar D^0 \to f} - \hat \Gamma_{D^0 \to f} \right) ~,
\end{equation}
\begin{equation} \label{eq:DYf}
\eta_f^{\rm CP} S_f = \eta_f^{\rm CP} 2 \Delta Y_f = x \left(\left|\frac{q}{p}\right| + \left|\frac{p}{q}\right|\right)\sin\phi - y \left( \left|\frac{q}{p}\right| - \left|\frac{p}{q}\right| \right) \cos\phi ~.
\end{equation}
Here $\eta_f^{\rm CP}$ is the CP parity of the final state $f$.
While singly Cabibbo suppressed decay modes can in principle be affected by new weak phases in the decay~\cite{Grossman:2006jg}, possible effects in the lifetime CP asymmetry are strongly constrained by existing data on time integrated CP asymmetries~\cite{Kagan:2009gb,Dmix_exp} and Eq.~(\ref{eq:DYf}) still holds to an excellent approximation. I.e. $\eta_f^{\rm CP} S_f $ is universal for all final states and practically independent of direct CP violation in the decays.
In fact, time dependent CP asymmetries are currently determined from the singly Cabibbo suppressed $D^0 \to K^+ K^-$ and $D^0 \to \pi^+ \pi^-$ modes and one has~\cite{Asner:2010qj}
\begin{equation}\label{eq:DYf_bound}
\eta_f^{\rm CP} S_f = (-0.246 \pm 0.496)\%~.
\end{equation}
Concerning Cabibbo favored decay modes, the most promising channel seems to be $D^0 \to K_S \phi$~\cite{Bigi:2009df}.

\begin{figure} \centering
\psfig{figure=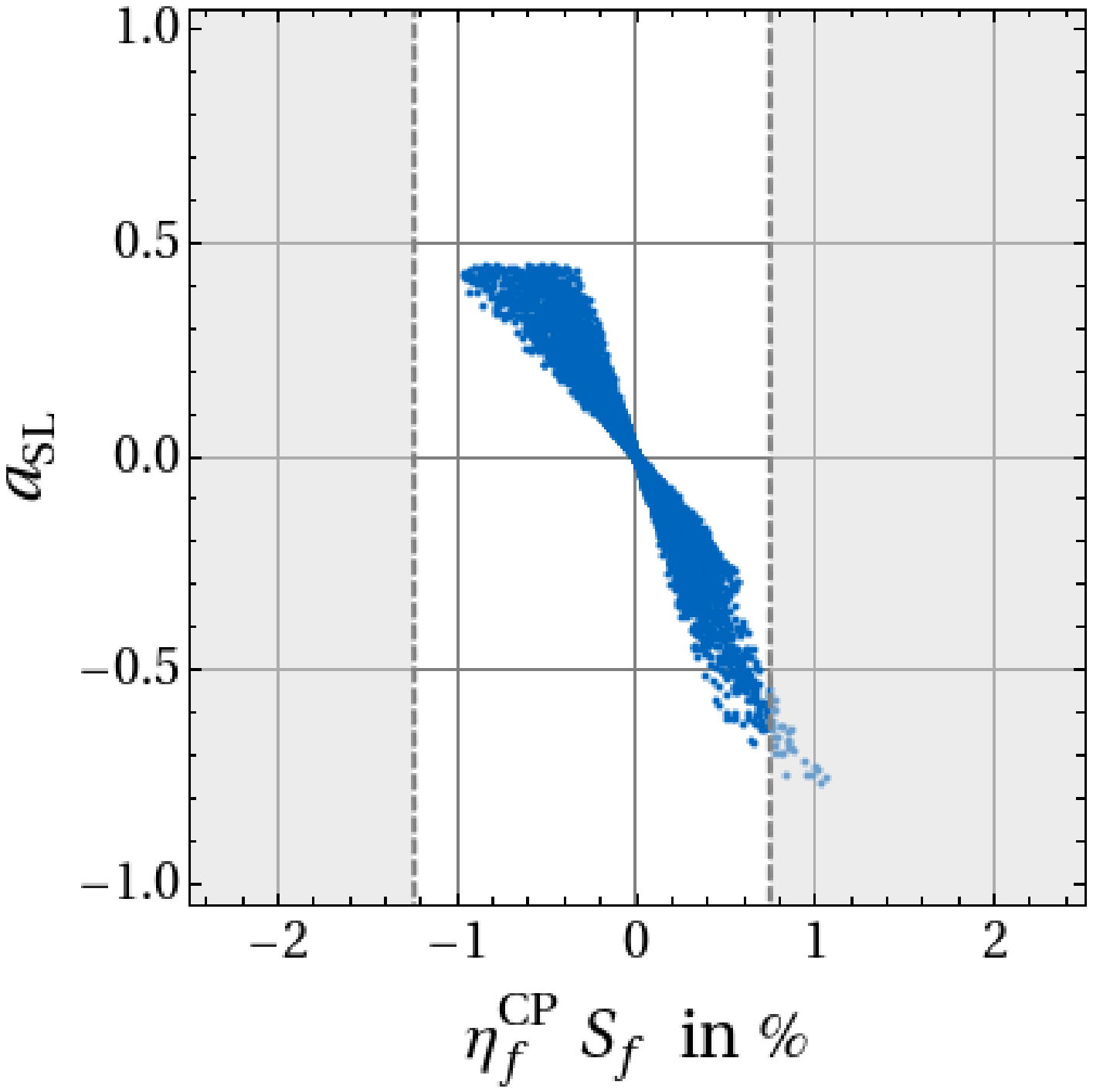,height=2.2in} ~~~~~~~~
\psfig{figure=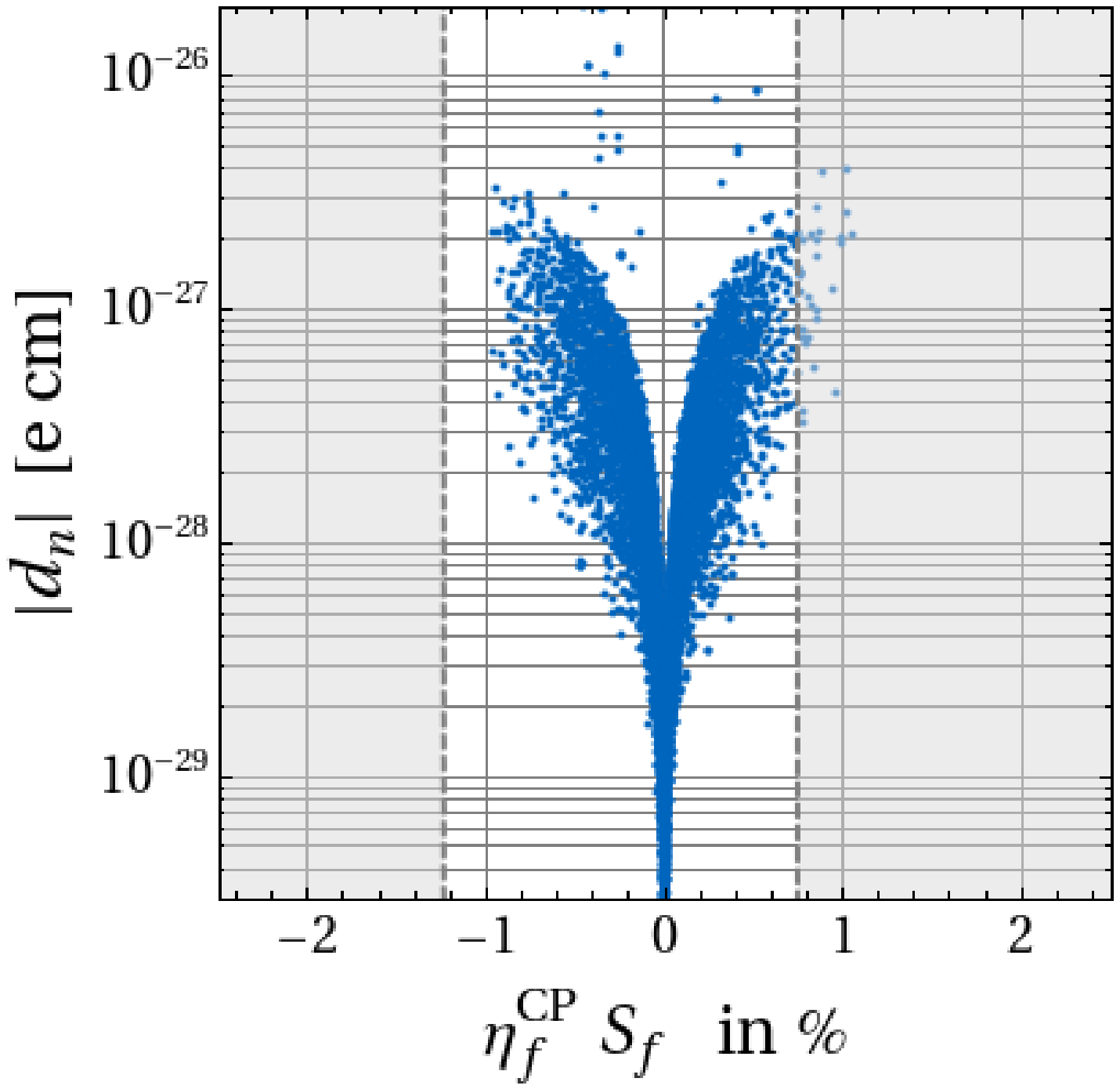,height=2.2in}
\caption{The semileptonic asymmetry $a_{SL}$ (left) and the neutron EDM $d_n$ (right) as a function of $S_f$ in SUSY alignment models. The gray region is excluded by the present data on $S_f$.
\label{fig:CPV}}
\end{figure}

In the left plot of Fig.~\ref{fig:CPV} we show the model independent correlation between $a_{SL}$ and $S_f$~\cite{Grossman:2009mn} in the context of SUSY alignment models. We assume a MSUGRA-like spectrum and scan the input parameter $m_0 < 2$TeV, $M_{1/2} < 1$TeV, $|A_0| < 3m_0$ and $5 < \tan\beta < 55$. At the GUT scale we fix $|(\delta_u^{RR})_{12}| = \lambda^3$ with an O(1) phase and set a mass splitting between the $1^{\rm st}$ and $2^{\rm nd}$ generation of squarks such that $m_{\tilde u_L} = 2m_{\tilde c_L} = 2m_0$. We find that in this setup the full range of values for $a_{SL}$ and $S_f$ that is compatible with the experimental constraints~(\ref{eq:Dmix_bounds}) can be reached.

\section{A Lower Bound on Hadronic EDMs in SUSY Alignment Models} \label{sec:EDM}

Electric Dipole Moments represent very clean probes of CP violation in extensions of the SM~\cite{Pospelov:2005pr}. While the SM predicts EDMs far below the present experimental bounds~\cite{EDMs_exp}
\begin{eqnarray} \label{eq:dTl_exp}
d_{\rm Tl} &\leq& 9.4 \times 10^{-25} ~e\,{\rm cm}~~~@~90\%~ \textnormal{C.L.}~, \\ \label{eq:dHg_exp}
d_{\rm Hg} &\leq& 3.1 \times 10^{-29} ~e\,{\rm cm}~~~@~95\%~ \textnormal{C.L.}~, \\ \label{eq:dn_exp}
d_{n}      &\leq& 2.9 \times 10^{-26} ~e\,{\rm cm}~~~@~90\%~ \textnormal{C.L.}~,
\end{eqnarray}
New Physics models that introduce new sources of CP violation are often strongly constrained by these bounds. In particular in the MSSM with SUSY particles at the TeV scale, flavor diagonal CP violating phases of e.g. the gaugino masses, the higgsino mass or the trilinear couplings are strongly constrained~\cite{EDMs_SUSY} at the level of $10^{-2}$.

In the MSSM with flavor violating soft terms, large NP effects for the hadronic EDMs can be naturally generated (see e.g.~\cite{Hisano:2008hn}). In particular, within SUSY alignment models, we find that the dominant SUSY contributions to the hadronic EDMs arise from ``flavored''  gluino -- up squark contributions to the up quark (C)EDM. At the SUSY scale one has
\begin{equation} \label{eq:edm_u_gluino}
\left\{ \frac{d_u}{e},~d^c_u \right\} \simeq -\frac{\alpha_s}{4\pi} ~m_c~ \frac{M_{\tilde g} A_c}{\tilde m^4_Q} ~\Big\{ f(x_g),~f^c(x_g) \Big\} ~{\rm Im} \left[(\delta_u^{LL})^*_{12}(\delta_u^{RR})_{12}\right] ~,
\end{equation}
with the loop functions $f$ and $f^c$ given e.g. in~\cite{Hisano:2008hn}. Even though this contribution is suppressed by a double flavor flip, the corresponding up quark (C)EDM is sizable due to the chiral enhancement by the charm quark mass. As in alignment models $(\delta_u^{LL})_{12}$ is real to an excellent approximation, the up quark (C)EDM~(\ref{eq:edm_u_gluino}) and CP violation in $D^0 - \bar D^0$ mixing~(\ref{ImMD_12}) is induced by the same combination of mass insertions. As also the charm trilinear coupling $A_c$ that enters~(\ref{eq:edm_u_gluino}) is naturally of the order of the gluino and squark masses, CP violating contributions to $D^0 - \bar D^0$ mixing automatically also imply a non-zero up quark (C)EDM that in turn will induce EDMs of hadronic systems like the neutron EDM $d_n$ or the mercury EDM $d_{\rm Hg}$, but not of the Thallium EDM $d_{\rm Tl}$.

In the right plot of Fig.~\ref{fig:CPV} we show the correlation between the time dependent CP asymmetry $S_f$ and the neutron EDM $d_n$ in SUSY alignment models (we use the same setup as described at the end of Sec.~\ref{sec:CPV_Dmix}). We observe that visible CP violating effects in $D^0 - \bar D^0$ mixing imply a lower bound on the neutron EDM. For $|S_f| > 0.1\%$ we find $d_n > {\rm ~few~} \cdot 10^{-29} \,e\,$cm and simultaneously for the mercury EDM $d_{\rm Hg} > {\rm ~few~} \cdot 10^{-31} \,e\,$cm which is an interesting level in view of future experimental sensitivities.

\section{Conclusions}

Electric Dipole Moments and CP violation in $D^0 - \bar D^0$ mixing are examples of low energy observables that are highly suppressed in the SM. Experimental evidence for them significantly above the tiny SM predictions would unambiguously signal the presence of NP.

Supersymmetric alignment models generically predict large non-standard effects in $D^0 - \bar D^0$ mixing~\cite{Nir:1993mx}.
In addition, as we demonstrated in~\cite{Altmannshofer:2010ad}, large CP violating effects in $D^0 - \bar D^0$ mixing in SUSY alignment models, generically also imply lower bounds for the EDMs of hadronic systems, like the neutron EDM and the mercury EDM, within the future experimental sensitivities. Correspondingly, the simultaneous evidence of CP violation in the neutral $D$ meson system together with non-vanishing hadronic EDMs would strongly support the idea of SUSY alignment models.

\section*{Acknowledgments}

I would like to thank the organizers for the invitation to Moriond EW 2011, Andrzej Buras and Paride Paradisi for the interesting collaboration and Stefania Gori for a reading of the manuscript. Fermilab is operated by Fermi Research Alliance, LLC under Contract No. De-AC02-07CH11359 with the United States Department of Energy.

\section*{References}

\end{document}